\documentclass[12pt]{article}

\newcommand{\om}{\omega}
\newcommand{\K}{\mathrm{K}}
\newcommand{\E}{\mathrm{E}}

\usepackage{graphicx}
\usepackage{amsmath}
\usepackage{amssymb}

\begin{document}
\title{Formation of soliton trains in Bose-Einstein condensates as a
nonlinear Fresnel diffraction of matter waves}

\author{A.M. Kamchatnov$^{1}$,
A. Gammal$^{2,3}$,
F.Kh. Abdullaev$^4$,\\ and
R.A. Kraenkel$^5$\\
$^1$Institute of Spectroscopy, Russian Academy of Sciences,\\ Troitsk 142190,
Moscow Region, Russia\\
$^2$Instituto de F\'{\i}sica, Universidade de S\~{a}o Paulo,\\
05315-970, C.P.66318 S\~{a}o Paulo, Brazil\\
$^3$Department of Physics and Astronomy and Rice Quantum Institute,\\
Rice University, Houston, Texas 77251\\
$^4$Physical-Technical Institute, Uzbek Academy of Sciences,\\ 70084 Tashkent-84,
G.Mavlyanov str. 2-b, Uzbekistan\\
$^5$Instituto de F\'{\i}sica Te\'{o}rica, Universidade Estadual Paulista-UNESP\\
Rua Pamplona 145, 01405-900 S\~{a}o Paulo, Brazil}

\date{\today}

\maketitle

%\pacs{03.75.Kk}

\begin{abstract}
The problem of generation of atomic soliton trains in elongated Bose-Einstein
condensates is considered in framework of Whitham theory of
modulations of nonlinear waves. Complete analytical solution is
presented for the case when the initial density distribution has
sharp enough boundaries. In this case the process of soliton train
formation can be viewed as a nonlinear  Fresnel diffraction
of matter waves. Theoretical predictions are compared with
results of numerical simulations of one- and three-dimensional
Gross-Pitaevskii equation and with experimental data
on formation of Bose-Einstein bright solitons in cigar-shaped traps.

\end{abstract}

Discovery of Bose-Einstein condensate (BEC) \cite{bec1,bec2,bec3} has created
new active field of research of quantum macroscopical behavior of
matter. Among most spectacular evidences of such macroscopic
behavior one can mention formation of interference fringes
between two  condensates \cite{andrews} and creation of
dark \cite{burger,denshlag} and bright \cite{khaykovich,strecker}
solitons. The interference
phenomenon is usually considered in framework of a linear wave
theory, whereas solitons are treated as a nonlinear wave effect.
At the same time, basically, these two phenomena have much in
common. For example, formation of bright soliton trains in
nonlinear wave systems is often explained  as a result of
modulational instability, where selection of the most unstable
mode is a result of interplay of interference and nonlinear
effects (see, e.g. \cite{fatkhulla,kamch2000}).
Such interconnection of interference and soliton
phenomena is demonstrated most spectacularly in formation of
solitons in vicinity of a sharp edge of density distribution. In
this case, at linear stage of evolution the linear diffraction
provides an initial modulation of the wave and further combined
action of interference and nonlinear effects leads to formation of
soliton trains. Without nonlinear effects, such kind of time
evolution of a sharp wave front would be a temporal counterpart of
usual spatial Fresnel diffraction and therefore soliton train
formation at the sharp front of nonlinear wave can be called a
nonlinear Fresnel diffraction.

Similar formation of oscillatory structures at sharp wave front or
after wave breaking in modulationally stable systems described by
the Korteweg-de Vries equation is well known
as a ``dissipationless shock wave'' (see, e.g. \cite{kamch2000}).
Its theoretical description is given \cite{gp,novikov} in
framework of Whitham theory
of nonlinear wave modulations \cite{whitham}, where the oscillatory structure
is presented as a modulated nonlinear periodic wave which parameters
change little in one wavelength and one period. Then slow
evolution of the parameters of the wave is governed by Whitham
equations obtained by averaging of initial nonlinear wave
equations over fast oscillations of the wave. Application of
this method to modulationally unstable systems has been given
for important particular case of soliton train formation at
the sharp front of a long step-like initial pulse \cite{k92,el,bk,k97}.
Here we shall consider by this method formation of solitons
in BEC with negative scattering length (attractive interaction
of atoms).

We suppose that condensate is confined in a very elongated
cigar-shaped trap whose axial frequency $\om_z$ is much less
than the radial frequency $\om_\bot$.
In the first approximation we can neglect the axial trap
potential and suppose that condensate is contained in a
cylindrical trap ($\om_z=0$) and its initial density
distribution has a rectangular form. Evolution of BEC is
governed by three-dimensional (3D) Gross-Pitaevskii (GP)
equation
\begin{equation}\label{GP}
i\hbar\psi_t=-\frac{\hbar^2}{2m_a}\Delta\psi+\frac12m_a\om_\bot^2
(x^2+y^2)\psi+g|\psi|^2\psi,
\end{equation}
for the condensate wave function $\psi$, where we use standard notation
$
%\begin{equation}\label{g}
g=4\pi\hbar^2a_s/m_a
%\end{equation}
$
for the effective nonlinear coupling constant,
$a_s<0$ is the $s$-wave scattering length,
and $\psi$ is normalized on the number of particles in BEC,
$
%\begin{equation}\label{norm3D}
\int|\psi|^2d\mathbf{r}=N.
%\end{equation}
$
For analytical treatment of nonlinear Fresnel diffraction it
is important to determine conditions when the 3D equation
(\ref{GP}) can be reduced to its one-dimensional (1D)
approximation (see, e.g. \cite{perez})
\begin{equation}\label{1D_GP}
i\hbar{\Psi}_{t}= -\frac{\hbar^2}{2m_a}\Psi_{zz}
+ g_{1D}|\Psi|^{2}\Psi,\quad \int|\Psi|^2dz=N,
\end{equation}
where
\begin{equation}\label{g1D}
g_{1D}=\frac{g}{2\pi a_\bot^2}=\frac{2\hbar^2a_s}{m_aa_\bot^2},
\quad a_\bot=\sqrt{\frac{\hbar}{m_a\om_\bot}},
\end{equation}
that is the transversal degrees of freedom are frozen.
It is well known (see, e.g. \cite{kamch2000}) that a homogeneous distribution
with linear density $n_0=|\Psi|^2=\mathrm{const}$ described
by (\ref{1D_GP}) with negative $g_{1D}$ ($a_s<0$) is unstable
with respect to self-modulation with increment of instability
equal in our present notation to
\begin{equation}\label{Gamma}
\Gamma=\frac{\hbar K}{2m_aa_\bot}\sqrt{8|a_s|n_0-({a_\bot K}
)^2},
\end{equation}
where $K$ is a wavenumber of small periodic modulation. The most
unstable mode has the wavenumber
\begin{equation}\label{Kmax}
K_{max}=2\sqrt{|a_s|n_0}/a_\bot
\end{equation}
and the corresponding increment is equal to
\begin{equation}\label{Gmax}
\Gamma_{max}=4|a_s|n_0\om_\bot.
\end{equation}
This means that after time $\sim1/(|a_s|n_0\om_\bot)$ the homogeneous
condensate splits into separate solitons (diffraction fringes) and
each soliton (diffraction fringe) contains about $N_s\sim n_0/K_{max}$
atoms. If in 3D GP equation (\ref{GP}) the nonlinear energy
$gN_sK_{max}/a_\bot^2\sim gn_0/a_\bot^2$ in each
solitons is much less than the kinetic energy in the transverse direction,
$\sim\hbar^2/m_aa_\bot^2$, then the transverse motion is reduced
to the ground state oscillations and the 3D condensate wave function
can be factorized into $\psi=\phi_0(x,y)\Psi(z,t)$, where
$\phi_0=(\sqrt{\pi}a_\bot)^{-1}\exp[-(x^2+y^2)/(2a_\bot^2)]$
is the ground state wave function of transverse motion, and $\Psi(z,t)$
obeys to the effective 1D nonlinear Schr\"{o}dinger (NLS) equation
(\ref{1D_GP}). Thus, the condition of applicability of 1D equation
(\ref{1D_GP}) for description of solitons formation is
\begin{equation}\label{cond}
n_0|a_s|\ll1,
\end{equation}
which means that the instability wavelength $\sim1/K_{max}$ is much
greater than the transverse radius $a_\bot$ of BEC.
%For typical value of the scattering length $|a_s|\sim 10^{-9}-10^{-8}\,$m
%this condition is fulfilled, if $N\sim 10^4$ and initial length of the
%condensate is about $L\sim 300\, \mu$m, so that $n_0=N/L\sim
%3\cdot 10^{7}\,\mathrm{m}^{-1}$. For greater values of $n_0$
If (\ref{cond}) is not satisfied, then the transverse motion
has to be taken into account which may lead
to collapse of BEC inside each separate soliton. Therefore we
shall confine ourselves to the BEC described by the 1D NLS
equation under supposition that the initial distribution
satisfies the condition (\ref{cond}).

To simplify formulae in the analytic theory, we transform (\ref{1D_GP}) to
dimensionless variables
$\tau=2(|a_s|n_0)^2\om_\bot t,$ $\zeta = 2|a_s|n_0{z}/{a_{\bot}},$
$\Psi = {\sqrt{2|a_{s}|}} n_0 u,$
so that (\ref{1D_GP}) takes the form
\begin{equation}\label{NLS}
iu_{\tau} + u_{\zeta\zeta} + 2|u|^{2}u =0,
\end{equation}
and $u$ is normalized to the effective length $L$ of the
condensate
$
\int|u|^2d\zeta=L/a_\bot
$
measured in units of $a_\bot$.
We are interested in the process of formation of solitons (nonlinear
Fresnel diffraction fringes) at the sharp boundary of initially
rectangular distribution. Since this process takes place
symmetrically at both sides of the rectangular distribution, we
can confine ourselves to the study of only one boundary. This
limitation remains correct until the nonlinear waves propagating
inside the condensate collide in its center. If the initial
distribution is long enough, this time is much greater than
the time of solitons formation. Thus, we consider the initial
distribution in the form
\begin{equation}\label{init}
u(\zeta,0) =\Bigg\{
\begin{array}{l}
\gamma \exp(-2i\alpha \zeta),\quad\mathrm{for}\quad \zeta <0 \\
   0, \quad\mathrm{for}\quad \zeta > 0,
\end{array}
\end{equation}
where $\gamma$ is the height of initial step-like distribution and
$\alpha$ characterizes the initial homogeneous phase.
The problem of this kind has already been considered in some other
problems of nonlinear physics \cite{k92,el,bk,k97,kamch2000} and we
shall present here only the main results.

Due to dispersion effects described by the second term in Eq.~(\ref{NLS}),
the sharp front transforms into slightly modulated wave which describes
usual Fresnel diffraction of atoms. In our case the diffraction pattern
evolves with time rather than is ``projected'' on the observation plane.
The linear stage of evolution is followed by the nonlinear one in which
combined action of dispersion and nonlinear terms yields the pattern
which can be represented as a modulated nonlinear periodic wave or,
in other words, a soliton train. We suppose that this soliton train
contains large enough number of solitons,
so that their parameters change little
in one wavelength and one period. Then, in framework of Whitham theory,
the density of BEC can be approximated by a modulated periodic solution
of Eq.~(\ref{NLS}) (see \cite{k97,kamch2000})
\begin{equation}
\label{n}
n = |u(\zeta,\tau)|^2 = (\gamma + \delta )^{2}
 - 4\gamma\delta\,\mbox{sn}^{2}
(\sqrt{(\alpha -\beta)^{2}
 + (\gamma + \delta)^{2}}\,\theta, m),
\end{equation}
where $\mbox{sn}(x,m)$ is the Jacobi elliptic function,
\begin{eqnarray}
\label{xi}
\theta = \zeta-V\tau,\quad V = -2(\alpha + \beta), \\
\label{m}
 m = {4\gamma \delta}/[{(\alpha -
\beta )^{2} + (\gamma + \delta )^{2}}],
\end{eqnarray}
the parameters $\alpha$ and $\gamma$ are determined by the initial
condition (\ref{init}), and $ \beta$ and $ \delta$ are slow functions
of $\zeta$
and $\tau$. Their evolution is governed by the Whitham equation
\begin{equation}\label{whitham}
\frac{\partial (\beta+i\delta)}{\partial \tau} + v(\beta,\delta)
\frac{\partial(\beta+i\delta)}{\partial \zeta} =0,
\end{equation}
where Whitham velocity $v(\beta,\delta)$ is given by the expression
\begin{equation}\label{v}
v(\beta,\delta)=-2(\alpha+\beta)\\
 -\frac{4\delta[\gamma-\delta+
i(\beta - \alpha )]\K}
{(\beta -\alpha )
(\K-\E) + i[(\delta -\gamma )\K +
(\delta + \gamma )\E]},
\end{equation}
$\K=\K(m)$ and $\E=\E(m)$ being the complete elliptic integrals of the
first and second kind, respectively. Since our initial condition
(\ref{init}) does not contain any parameters with dimension of length,
the parameters $\beta$ and $\delta$
can only depend on the self-similar variable $\xi = \zeta/\tau$.
Then Eq.~(\ref{whitham}) has the solution
\begin{equation}\label{sol}
{\zeta}/\tau= \xi = v(\beta,\delta)
\end{equation}
with $v(\beta,\delta)$ given by (\ref{v}). Separation of real and imaginary
parts yields the formulae
\begin{eqnarray}
\label{imp1}
{\zeta}/{\tau} =-4\beta-{2(\gamma^2-\delta^2)}/({\beta-\alpha}),\\
\frac{(\alpha-\beta)^2+(\gamma-\delta)^2}{(\alpha-\beta)^2+\gamma^2-
\delta^2}=\frac{\E(m)}{\K(m)},
\end{eqnarray}
which together with Eq.~(\ref{m}) determine implicitly dependence of
$\beta$ and $\delta$ on $\xi=\zeta/\tau$.
It is convenient to express this dependence in parametric form
\begin{eqnarray}
\label{par1}
\beta(m) = \alpha - \gamma\sqrt{4A(m) - (1+m A(m))^{2}}, \\
\label{par2}
 \delta(m) = \gamma m A(m),
\end{eqnarray}
where
\begin{equation}
A(m) = \frac{{(2-m)\E(m) - 2(1-m)\K(m)}}
{m^{2}\E(m)}.
\end{equation}
Substitution of these expressions into (\ref{n}),(\ref{xi})
yields the density $n$ as a function of $m$. Since the space
coordinate $\zeta$ defined by Eq.~(\ref{imp1}) is also a function of
$m$ at given moment $\tau$, we arrive at presentation of dependence
of $n$ on $\zeta$ in parametric form. The limit $m\to0$ corresponds to
a vanishing modulation, and this edge point moves inside the
condensate according to the law
\begin{equation}\label{zminus}
\zeta_{-} = (-4\alpha + 4\sqrt{2}\gamma)\tau.
\end{equation}
The other edge with $m\to1$ moves according to the law
\begin{equation}\label{zplus}
\zeta_{+} = -4\alpha \tau,
\end{equation}
and corresponds to the bright solitons (or fringes of nonlinear diffraction
pattern) at the moment $\tau$. The whole region $\zeta_-<\zeta<\zeta_+$
describes
the oscillatory pattern arising due to nonlinear Fresnel diffraction
of the BEC with initially sharp boundary at $\zeta=0$.

We have performed numerical simulation of 1D and 3D GP
equations with the aim to compare approximate Whitham theory with
numerical results. The 1D density distributions calculated
numerically from (\ref{NLS}) and analytically are shown in Fig.~1.
We see
excellent agreement between the theoretical and numerical
predictions of the height of the first soliton generated
from initially step-like pulse, but its position given by
analytical formula is slightly shifted with respect to
numerical result. This is well-known feature of asymptotic
Whitham approach \cite{gp,novikov} which accuracy in
prediction of location of the oscillatory pattern cannot be
much better than one wavelength. Thus, we see that the
above theory reproduces the numerical results quite well
for period of time $\tau\simeq 2$. For much greater time
values some other unstable modes different from one-phase
periodic solution (\ref{n}) can also give considerable
contribution into wave pattern. Nevertheless, the
qualitative picture of soliton pattern remains the same.

For 3D numerical simulation, the GP equation (\ref{GP}) was
transformed to dimensionless form by means of substitutions
$x=a_\bot x',$ $y=a_\bot y',$ $z=a_\bot z',$ $t=2t'/\om_\bot,$
$\psi'=(N^{1/2}/a_\bot^{3/2})\psi,$ so that it takes the form
\begin{equation}\label{GP2}
i\psi_t=-\Delta\psi+r^2\psi-(8\pi N|a_s|/a_\bot)|\psi|^2\psi,
\end{equation}
where  primes are omitted for
convenience of the notation and
$\int|\psi|^22\pi rdrdz=1$, $r^2=x^2+y^2$.
Evolution of the density distribution
$\rho(z)=\int_0^{\infty}|\psi(r,z)|^22\pi rdr$ along the axial
direction is shown in Fig.~2 for the values of the parameters
corresponding to the experiment \cite{strecker} ($a_s=-3a_0,$
$\om_\bot=2\pi\cdot625\,\mathrm{Hz},$ $L=300\,a_\bot$) except
for the number of atoms which was chosen to be $N=5\cdot 10^3$
in order to satisfy the condition (\ref{cond}), so that
$|a_s|n_0=1.7\cdot 10^{-3}$.
We see that diffraction (soliton) pattern arises after the
dimensionless time $t\simeq400$ which corresponds after appropriate
scaling transformation to $\tau\simeq2$ in Fig.~1. The width of
solitons in Fig.~2 also agrees with the width
predicted by 1D analytical theory and numerics. The spatial
distribution of the condensate density $|\psi(r,z)|^2$ is
illustrated by Fig.~3.
The 3D nonlinear interference pattern is clearly seen. For greater
values of the condensate density, when 1D theory does not apply, numerical
simulation demonstrates similar evolution of the diffraction pattern
up to the moment when collapse starts in each separate soliton.
Thus, formation of solitons in the experiment \cite{strecker}
with large initial number of atoms $N\simeq10^5$ goes through
collapses with loss of atoms until the remaining atoms can form
stable separate soliton-like condensates. The present theory
emphasizes the importance of the initial stage of evolution with
formation of the nonlinear Fresnel diffraction pattern.

Formation of soliton trains in BEC confined in a cigar-shaped trap
has also been studied numerically in \cite{carr,salasnich}.
The results of 1D simulation in \cite{carr} agree qualitatively
with our results. In numerics of  \cite{salasnich} strong losses
were introduces to prevent fast collapse of BEC with large number
of atoms. Nevertheless, formation of soliton trains was also
observed.

The above theory is correct for evolution time much less than
period of oscillations $2\pi/\om_z$ in the axial trap. When
the axial trap is taken into account, solitons acquire velocities
in axial direction even if initial phase is equal to
zero. The number of solitons produced ultimately from some finite
initial BEC
distribution can be found by means of quasiclassical method applied
to an auxiliary spectral problem associated with the NLS equation
(\ref{NLS}) in framework of the inverse scattering transform method
\cite{novikov,kku02}.
If the initial wave function
is represented in the form $u_0(\zeta)=\sqrt{n_0(\zeta)}
\exp(i\phi_0(\zeta))$,
then the total number of solitons is equal approximately to
\begin{equation}\label{n_s}
N_s=\frac1{\pi}\int\sqrt{n_0(\zeta)+\frac{v_0^2(\zeta)}4}d\zeta
-\frac12,
\end{equation}
where $v_0(\zeta)=\partial\phi_0(\zeta)/\partial\zeta$ is the
initial velocity distribution of BEC.
If there is no
initial phase imprinted in BEC, then the total number of solitons
is given by the formula
\begin{equation}\label{Nsol}
N_s=({\sqrt{2|a_s|}}/{\pi a_\bot})
\int|\Psi|dz,
\end{equation}
which is written in dimensional units
and we have neglected a ``one-half'' term in (\ref{n_s}).

In experiment,
the initial stage is usually obtained by sudden change of the sign
of the scattering length from positive to negative one, so that
initial density distribution has, for large enough number of atoms,
the Thomas-Fermi form
\begin{equation}\label{TF}
|\Psi|^2=({3N}/{4Z})\left(1-{z^2}/{Z^2}\right),
\end{equation}
where $Z$ is the Thomas-Fermi half-length of the condensate.
Then substitution of
(\ref{TF}) into (\ref{Nsol}) gives
\begin{equation}\label{nsol2}
N_s=\sqrt{3N|a_s|L}/(4a_\bot),
\end{equation}
where $L=2Z$ is the total length of the condensate.  Up to constant factor,
this estimate coincides with
one obtained in \cite{salasnich} by division of $L$ by the instability
wavelength $1/K_{max}$. Note that this estimate includes also very
small solitons which cannot be observed in real experiments, so that it must
be considered as an upper limit of the number solitons which can be produced
from a given initial distribution.
The same property of this kind of estimate
for number of dark  solitons has been observed in comparison of
analytical theory with numerical simulations in \cite{kku02}. The axial
potential influences mainly on velocities of solitons, so the above
estimate can be applied to the condensate in a cigar-shape trap
under condition that inequality (\ref{cond}) is fulfilled.

In conclusion, we have studied theoretically and numerically the
process of formation of soliton trains near the sharp edges of
the density distribution of BEC. The arising oscillatory regions
can be considered as nonlinear Fresnel diffraction fringes
of matter waves.

This work was supported by FAPESP (Brazil) and CNPq (Brazil).
A.M.K. thanks also
RFBR (grant 01--01--00696) for partial support. A.G. would like to
thank Randy Hulet and his group for useful discussions.

\newpage

\begin{figure}[ht]
\centerline{\includegraphics{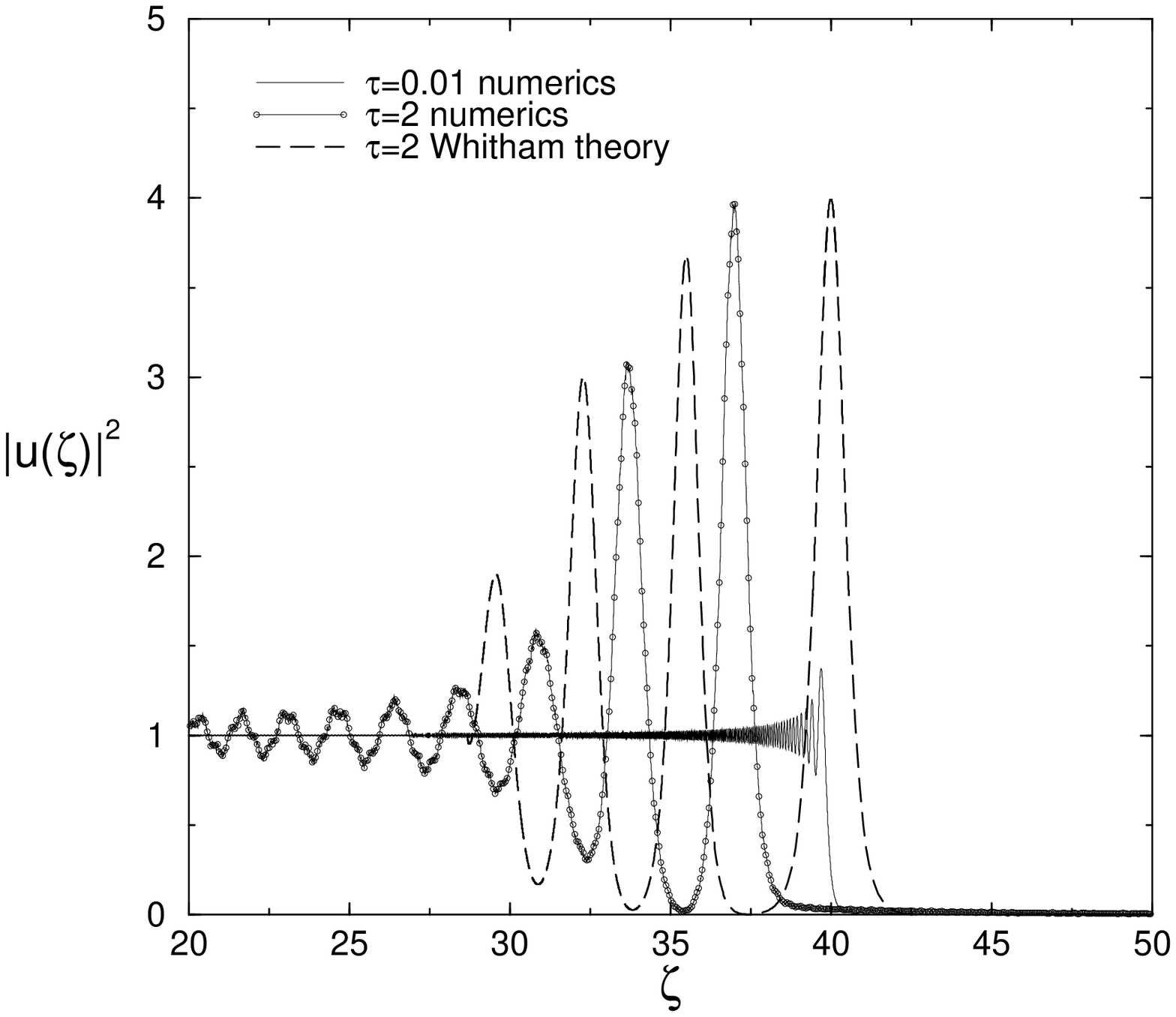}}
\vspace{0.3 true cm}
\caption{Density distributions of BEC  calculated
by numerical solution of 1D GP equation (\ref{NLS}) and given by
Whitham theory with initial step-like wave function (\ref{init})
with  $\gamma=-1,$  $\alpha=0$.
}
\label{figone}
\end{figure}

\newpage

\begin{figure}
\vspace{0.3 true cm}
\centerline{\includegraphics{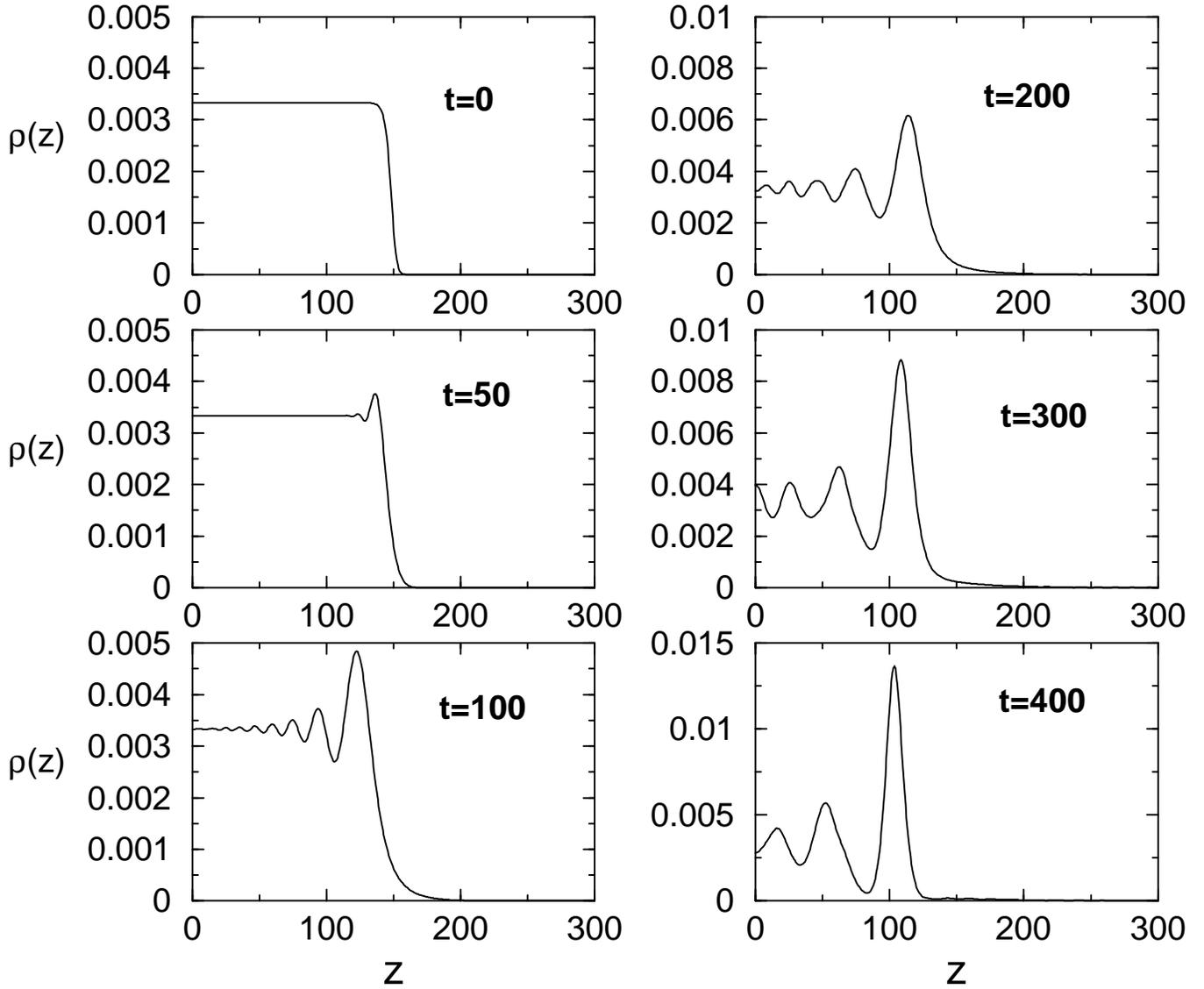}}
\caption{Density distributions of BEC $\rho(z)$ along axial direction
for different moments of time  calculated
by numerical solution of 3D GP equation (\ref{GP2}) with cylindrical
initial distribution.
}
\label{figtwo}
\end{figure}

\newpage

\begin{figure}[ht]
\centerline{\includegraphics{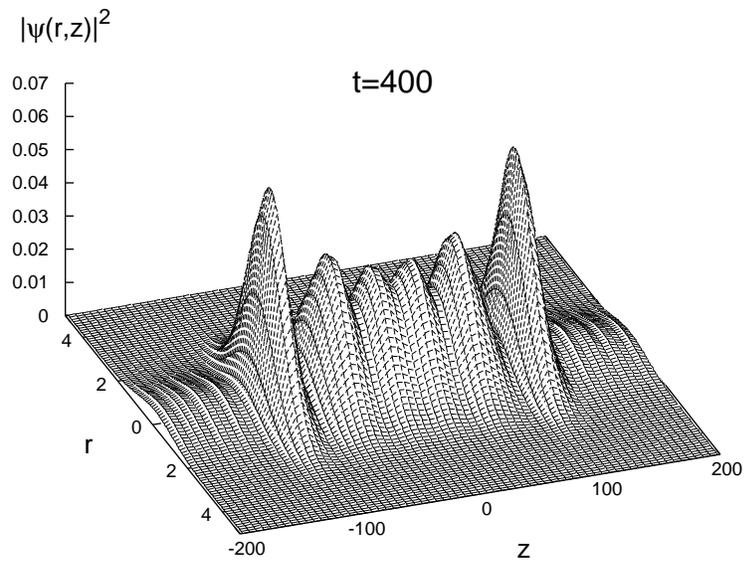}}
\vspace{0.3 true cm}
\caption{Dependence of the density distributions on radial, $r$,
and axial, $z$, coordinates at time $t=400$.
}
\label{figthree}
\end{figure}

\end{document}